\documentclass[aip,rsi,reprint]{revtex4-1} 

\usepackage{siunitx}
\usepackage{amsmath}
\usepackage{amssymb}
\usepackage{todonotes}

\begin{document}

\title{Antireflection Coated Semiconductor Laser Amplifier}

\author{Vasiliki Bolpasi}
\email[]{vasbol@iesl.forth.gr}
\affiliation{IESL - FORTH}

\author{Wolf von Klitzing}
\email[]{wvk@iesl.forth.gr}
\affiliation{IESL - FORTH}

\date{\today}
\begin{abstract}
This paper presents a laser amplifier based on antireflection coated diodes.
It operates without active temperature stabilisation at any wavelength within its gain profile without restrictions on the injection current.
Using a active feedback from an external detector, the power of the amplied light can be stabilized to better than $20\mu W$ or $2\times 10^{-4}$, even after additional optical elements such as  opticals fibers and/or polarizating beam splitters.
The power  of the amplied light  can be ramped and modulated  arbitrarily without loss of stability in the output power.
In the absence of the seeding light, the laser amplifier outputs little directed light. 
The construction of the laser amplifier is extremely simple, requiring neither active temperature stabilization nor a device for the external monitoring the spectral purity of the light.
\end{abstract}


\maketitle 

\section{Introduction}
\begin{figure}[t]
  \includegraphics[width=0.45\textwidth]{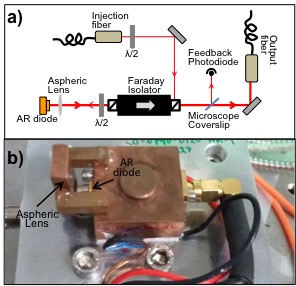}
  \caption{ a) Diagram of the optical setup and b) photo of the mount and collimation of the laser diode.}
   \label{fig:MountPhoto}
\end{figure}

Diode lasers offer good reliability and high power levels at a relatively low cost. Their application ranges from telecommunications to environmental monitoring and atomic physics. 
A high degree of spectral purity can be achieved via a combination of optical and electronic feedback \cite{Corwin1998AO,Wieman1991ROSI,Santarelli1994OC,Talvitie1998OC,Schwarze1999ROSI,Robins2002OL,Sahagun2013OC} even reaching line widths well below 1\,Hz, albeit at low output powers\cite{Stoehr2006OL,Kessler2012NP}. 
One can amplify the light using high-power tapered laser diodes reaching up to a few watts \cite{Shima1990QEIJO,Bolpasi2011P}.
If more moderate powers are required, then the light can be amplified by injection locking a standard ridge-guide laser diode, which is then commonly referred to as \emph{slave laser}.
Typically, this is done by injecting a small amount of light into the slave diode through the side-port of an optical isolator (see fig. \ref{fig:MountPhoto}).  
The  slave laser then amplifies and reflects the light, which then exits the optical isolator through its output port.

In standard laser diodes, the front facet has a simple anti-refection (AR) coating and the back side a highly reflective coating, with the two facets forming  a weak Fabry-Perot (FP) cavity. 

For slave lasers to operate at low injection powers, the optical length of the cavity has to match the wavelength of the light to be amplified.
Therefore, the slave diode can be locked only in narrow ranges in the injection current and in the temperature of the cavity \cite{Saxberg2016ROSI}.
In between these ranges, the diode will lase at its own resonance frequency.
As a result, the FP-diode based slave-lasers  can amplify only one wavelength at a time and its intensity cannot be modulated or continuously tuned or even stabilized using the injection current.

In this article, we present a novel type of \emph{slave laser}, which is based on a laser-diode with an output facet with multi-layer anti-reflection coating (AR-diode).  
Due to the absence of a laser-resonator this slave laser does not emit directed radiation other than the amplified seed light--thus eliminating the need for a dedicated analysis of the emitted light
\footnote{For FP-diode based slave lasers,  usually a fraction of the light is sent to an external FP-cavity in order to ensure that that the emitted light is exclusively amplified light and that no other lasing modes are present. Note that an FP-diode will lase even without seeding.}. 
Another consequence is that slave-laser is capable of amplifying multiple wavelengths simultaneously and can be continuously tuned and modulated in intensity. 
Furthermore, the optical length of the cavity does not play a significant role, thus eliminating the need for an active temperature stabilization.
With the need for an accurate temperature control and dedicated FP-cavity and electronics eliminated, the AR slave laser setup developed here consists only of a passively cooled laser diode, current driver, and an optical isolator. 

 \begin{figure}
   \includegraphics[width=0.45\textwidth]{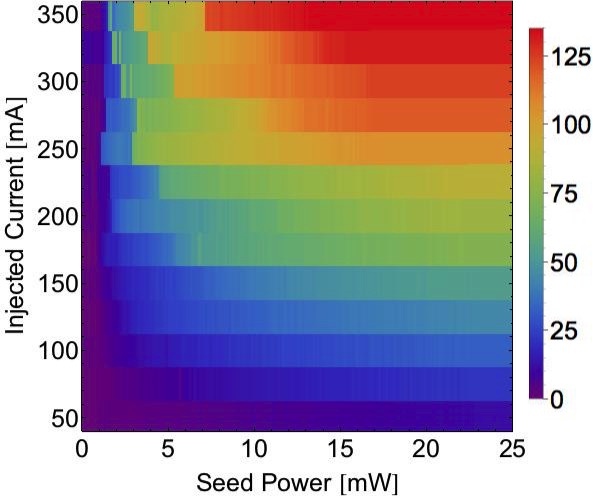}
   \caption{Dependence of the emitted power from the seeding power and the injection current. Note that the maximum output power of 125\,mW can be reached already for 7\,mW seed power.  The color scale of the emitted power is in units of milli watt. 
  }
   \label{fig:Seed}
\end{figure}

\section{Setup}

Fig.\,\ref{fig:MountPhoto} depicts a schematic diagram of the slave laser setup.
The diode under investigation is a 140\,mW, antireflection coated laser diode (Topitca LD-0790-0120-AR-1). 
The diode is mounted onto a  copper block and collimated by an aspheric lens (Thorlabs C230TM-B, f=4.5\,mm), which is bonded to the main block by two copper pillars using a standard epoxi adhesive. 

The temperature of the laser diode is controlled via a Peltier element, which is bonded on using a combined laser-current source and temperature controller (Stanford LDC501).
In order to collimate the laser beam, the lens was actively positioned using a copper lens mount and a 3D translation stage, and then bonded to the pillars using standard epoxy adhesives.
The injection beam is delivered from the master laser to the slave setup via a polarization-maintaining single-mode FC-APC optical fiber. 
After  collimation (Schafter-Kirchhoff 60fc-4-A8-07), it is injected into the laser diode via the auxiliary port of a 35 dB Faraday isolator (Electro-Optics Technology 4I780-MP).
The amplified beam passes then through the isolator and is coupled in a single mode optical fiber.
A microscope cover-slip samples a small portion of the light and sends it to a photo-diode for monitoring and feed-back purposes.

\section{Characterization of the diode}

  \begin{figure}
   \includegraphics[width=0.45\textwidth]{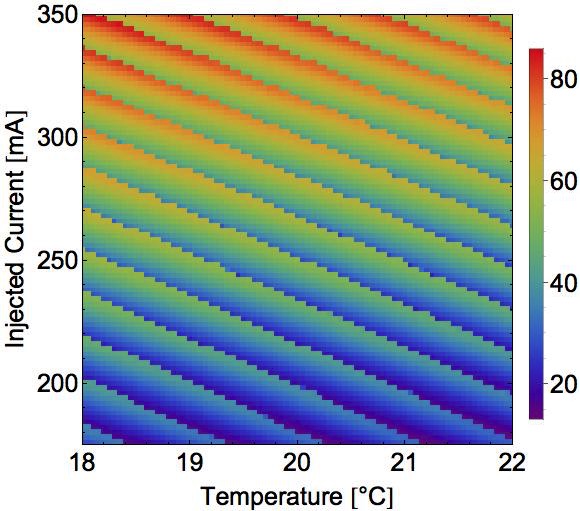}
   \caption{Plot of the measured output power vs the current and the temperature. The data was taken for an injection beam power of approximately 3\,mW. The color scale is in units of milli watt.}
   \label{fig:ColorPlot}
\end{figure}

\subsection{Lasing properties}

The main effect of the AR coating on the laser diode is to push the self-lasing threshold to values close to the maximum permitted injection current.
Below this threshold the laser diode emits mainly diffuse light. 
Just below the lasing threshold (350\,mA and $22^{\circ}$C) only 1\,mW is coupled into the fiber with its spectrum being broad without any major peaks. 
Above a critical injection current the diode starts to lase on its own and a peak appears at 767.1\,nm. 
When injected with the light from the master laser the free-running peak and the background disappear and the spectrum contains only a peak at the master's wavelength.  
For our laser diode the lasing threshold lies at 18$^{\circ}$C and 337\,mA. Above 19.1$^{\circ}$C we do not observe lasing for any current below the maximum specified levels (350\,mA).

A complication in FP-diode based slave lasers is that the output light may be amplified light or originate from self-lasing, depending on the laser current, temperature and seeding power.
This necessitates an external cavity to monitor the spectral purity of a FP-diode based laser amplifier (salve laser). 
The AR-diode does not suffer from this: Since the laser diode does not self-lase (e.g.\,above 19.1$^{\circ}$C), it prouces very little directed output, and any light coupled into the fiber is amplified light originating from the master laser.

\subsection{Output power vs seeding power}
In fig.\ref{fig:Seed} shows the  output power vs the seeding power and the injection current.
\footnote{Note that the seeding power is measured just before the slave laser and the output power just after the isolator.}
For a given injection current, one observes as a function of the power of the seed-light  a rapid increase of the output power which rapidly saturates. 
Note that the maximum output power can already be reached for a seed power of 7\,mW. 
For a given injection power the rise in output power is much more smooth.

 \begin{figure}
   \includegraphics[width=0.45\textwidth]{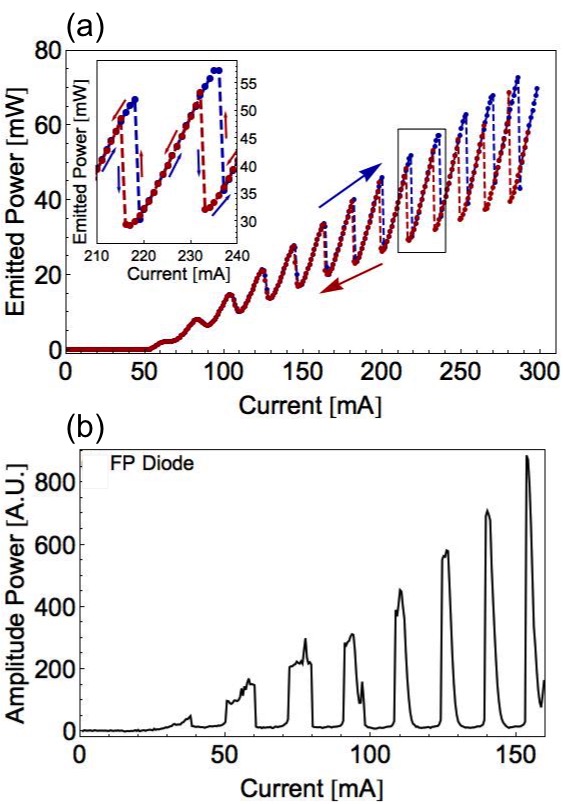}
   \caption{Ramping the current in constant current (CC) mode. a) The output power of the AR-coated slave diode for 3\,mW of seeding power at 20$^{\circ}$C. The power is ramped up (blue) and down (red). Note that amplification occurs at any current and power. b) The output power of a slave based on an FP-diode (Sharp GH0781JA2C) for \textcolor{blue}{1-3}\,mW of seeding power. Note that the amplification works only for narrow ranges of currents and temperatures. }
   \label{fig:CC}
\end{figure}

  \begin{figure}
   \includegraphics[width=0.45\textwidth]{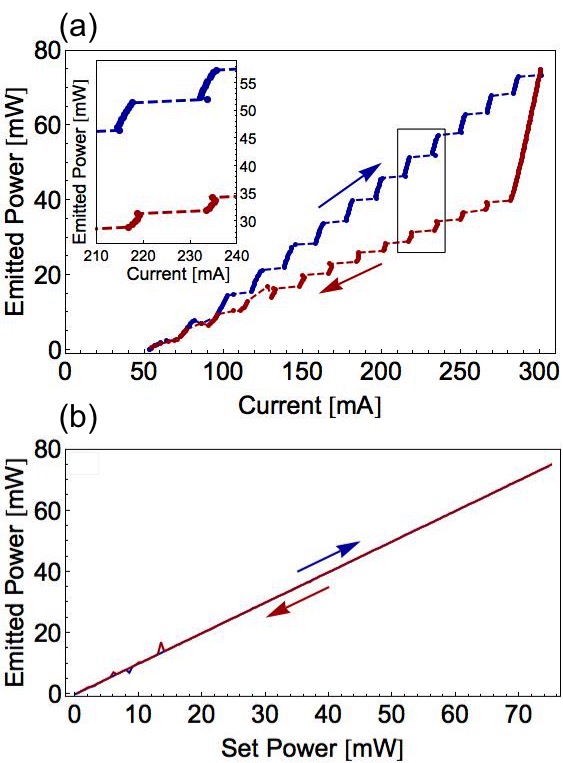}
   \caption{In constant power mode, the set power is ramped upwards and downwards again. In a) the measured output power is plotted against the current and in b) against the set power. Above the value of 14\,mW ,where the fluctuations are limited the standard deviation of the difference between the programmed values of the power and the actual emitted power, is only 56.6\,$\mu$W.}
   \label{fig:CP}
\end{figure}

 \subsection{Output power vs temperature and current} \label{sec:OutputPowerVsTandI}

A perfect AR-coating would completely remove the FP cavity effects. A change in temperature of the laser diode and/or its injection current would  only cause a very smooth change in the shape of the gain profile.  
For imperfect AR-coatings one observes FP fringes, which depend on the exact optical length of the cavity, which in turn changes rapidly with  temperature and  injection current. 
The resulting fast modulation on a slow background can be seen in Fig.\,\ref{fig:ColorPlot}, which shows a plot of the output power of the AR-diode vs current and  temperature.
As the current increases, the output power slowly rises to the maximum allowed value.
On top of this slow variation there is a very regular sawtooth-like modulation which depends both on the current and the temperature of the diode.

The data in fig.\,\ref{fig:ColorPlot} was taken in many cycles. For a given current value, the temperature was ramped up from 18$^{\circ}$C to 22$^{\circ}$C. This procedure was repeated for all current values, starting from 175\,mA up to 330\,mA. The step size for the temperature was 0.05$^{\circ}$C and for the current it was 2\,mA.

Fig.\ref{fig:CC}a  shows the power of the AR coated slave diode after the output fiber  versus the current. 
The expected slow increase of the output power with the current is clearly visible together with a fast saw-tooth-shaped modulation.
In the insert a clear hysteresis loop can be seen.  
The blue data points were taken for increasing current and the red ones for decreasing current

The origin of the hysteresis is not entirely clear, but is likely to originate from a combination of the the FP-cavity (formed by the back-facet of the diode and small reflection from its AR-coated facet) and the injection-current dependent refractive index within the cavity.
Jumps in the power for a small change in current could prevent for some currents a stable operation of the laser diode in the unavoidable presence of noise. 
However,  for sufficiently low noise the hysteresis loop guarantees stable operation at any current. 

\subsection{Power stabilization with feedback loop}  

Fig.\,\ref{fig:CC}b shows a  plot of the amplified light versus the injection current for a slave based on a FP-diode, which was discriminated from other lasing modes using a grating spectrometer. 
Clearly, for a given temperature, there are only narrow ranges of currents, where an amplification occurs.
This is due to the facets of the FP-diode creating an optical cavity. 
The resonant frequency of the cavity tunes with the injection current via the refractive index of the gain medium and the mechanical length of the cavity due to changes in temperature.
At certain current the cavity is resonant and amplification occurs; at other currents it is anti-resonant. In the latter case the seeding light cannot enter the cavity and the `slave-laser' will self-lase at a wavelength different from the one of the master laser.

Due to the lack of cavity in the AR-coated diode, there is no preferable mode that is amplified, and therefore no particular conditions need to be met (i.e.\,case temperature and laser current) for the diode to lock to the seeding light.
Therefore, as opposed to the FP-diode, the AR-diode can be operated at any current. 
One can therefore power-stabilise the AR-coated slave laser by measuring the power and feeding the result back to the current using a PID circuit.
A particularly interesting application of this is to stabilize the power not at the diode itself, but e.g.\,at the magneto-optical trap after transmission through an optical fiber and a polarizer.
Figure \ref{fig:CP}a shows the feed-back stabilized output power versus the \emph{current} as adjusted by the feedback loop, both for a sweep upwards and downwards.
Here, the jumps of the hysteresis loop are clearly visible.
However, they do not affect the stability of the output power as can be seen in fig.\,\ref{fig:CP}b, which shows the output power of the stabilized slave laser as a function of the programmed \emph{power}. 
Note that there are no gaps, nor jumps, nor any hysteresis in the data.

\begin{figure}
   \includegraphics[width=0.45\textwidth]{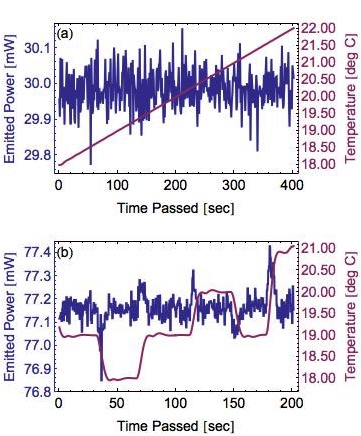}
  \caption{Power of the amplified light in constant power mode  (a) Ramping the temperature by 0.6\,K/min. (b) Applying temperature shocks of up to 1.5\,K/min. 
   The left axis of the graph (blue color) is the emitted power, and the right axis (purple color) is the corresponding case temperature. The stability is better than $50\,\mu \text{W}_\text{RMS}$. In the absence of large temperature fluctuations the laser is stable to $20\,\mu \text{W}_\text{RMS}$ or $2*10^{-4}$}
  \label{fig:P-t}
\end{figure}

Nevertheless, we do expect very short-lived jumps in the intensity as we ramp temperature or injection current.
To verify this, we examine the output power of the diode versus time (see fig.\,\ref{fig:P-t}), whilst forcing the AR diode to undergo big changes in temperature.
Figure\,\ref{fig:P-t}a shows the output power of the power-stabilized slave laser. The photodiode of the  feedback was placed after the optical fiber and a polarising beam splitter. 
We thus simulate an experimental situation, where one wants to stabilize the power at a particular point in the experiment after a variable lossy channel.
The laser diode was subjected to a smooth temperature ramp from 18$^{\circ}$C to 22$^{\circ}$C.
The measured RMS fluctuations of the power were only $50\,\mu \text{W}_\text{RMS}$ in 30\,mW.
In a more stringent test, we applied sudden jumps in temperature by 2\,K (see fig.\,\ref{fig:P-t}b). Despite these abrupt changes the power deviates by less than 0.2$\%$ and stabilizes to its original value within a few seconds.
Finally, in the absence of large temperature fluctuations of laser the power is stable to $20\,\mu \text{W}_\text{RMS}$ or $2*10^{-4}$.

\section{Conclusions}
We investigated a slave laser configuration based on a highly AR-coated laser diode, that provides up to 125\,{\rm mW} of useful output power for 7\,mW seed power. 
The setup is simplified with respect to the setups using conventional FP diode lasers:
It does not require active cooling or temperature stabilization neither--as opposed to traditional laser diode amplifiers-- monitoring of the spectral quality of the amplification.
Finally, the output power of the laser amplifier based on AR coated diodes can be modulated, continuously ramped, or even stabilised to better than 0.1\%.
Expected applications are in cold atom experiments and especially in the context of mobile and space-based quantum-enhanced sensors.

\begin{acknowledgments}
We acknowledge the financial support of the Future and Emerging Technologies (FET) programme within the 7th Framework Programme for Research of the European Commission, under FET grant number: FP7-ICT-601180 and of a Marie Curie Excellence grant under Contract MEXT-CT-2005-024854. 
We also acknowledge funding from the European Space Agency under contract No. 4000112744/14/NL/PA.
\end{acknowledgments}

\bibliography{laser}

\end{document}